\documentclass[12pt]{article}
\usepackage{graphicx} 
\usepackage{dcolumn}
\usepackage{bm}
\input{epsf.sty}
\usepackage{texdraw}

\newcommand{\gam}{\gamma}
\newcommand{\Gam}{\Gamma}

\newcommand{\omt}{{\omega_0t\,}}

\newcommand{\si}{\hat{\sigma}}

\newcommand{\om}{\omega}

\newcommand{\tr}{{\rm tr}}
\newcommand{\ha}{\hat{a}}

\newcommand{\BEQ}{\begin{equation}}
\newcommand{\EEQ}{\end{equation}}
\newcommand{\BEA}{\begin{eqnarray}}
\newcommand{\EEA}{\end{eqnarray}}
\newcommand{\nn}{\nonumber}

\renewcommand{\d}{{\rm d}}
\newcommand{\p}{\partial}
\newcommand{\eps}{\varepsilon}

\renewcommand{\H}{{\cal H}}
\newcommand{\RS}{{\rm S}}

\newcommand{\T}{{\cal T}}

\newcommand{\half}{\frac{1}{2}}
\renewcommand{\thesection}{\arabic{section}}

\def\dbarrm {{\mathchar'26\mkern-11mu{\rm d}}}                       %
\begin{document} 
\title{Thermodynamics and small quantum systems}
\author{Th.M. Nieuwenhuizen
\\Institute for Theoretical Physics, University of Amsterdam,\\
Valckenierstraat 65, 1018 XE Amsterdam, The Netherlands}
\maketitle

\begin{abstract}
Small quantum systems non-weakly coupled to a bath become 
in the quantum regime surrounded by a cloud of
photons or phonons, which modifies their thermodynamic behavior.
Exactly solvable examples are the Brownian motion of a quantum particle 
in a harmonic confining potential and coupled to a harmonic quantum 
thermal bath, e.g. an ion in a Penning trap, 
and a spin immersed in a bosonic bath, as occurs in NMR physics.
It appears that the Clausius inequality $\dbarrm Q\le T\d S$ can be violated.
For non-adiabatic changes of system parameters the rate of energy 
dissipation can be negative, and, out of equilibrium, cyclic processes 
are possible which extract work from the bath.
Experimental setups for testing some of the effects are discussed.
\end{abstract}

\renewcommand{\thesection}{\arabic{section}}
\section{ Introduction}
\setcounter{equation}{0}\setcounter{figure}{0} 
\renewcommand{\thesection}{\arabic{section}.}

The microscopic basis of thermodynamics is statistical physics and 
equilibrium is described by the Gibbs distribution. 
It is typically taken for granted that, when going to the quantum regime, 
the classical Gibbs distribution can just be replaced by its quantum analog.
It is not often stressed that this is only allowed in case of
weak coupling with the bath. 
When that coupling is not weak, the Gibbs state
of the total system (subsystem+bath) leads, after tracing out the bath,
to a non-Gibbsian state for the subsystem.
This endangers (near-) equilibrium thermodynamics. 
We analyze the situation for two exactly solvable problems.

A Letter on the thermodynamics of quantum Brownian motion
appeared~\cite{ANQBMprl}, and was discussed in the scientific literature
~\cite{AIP}~\cite{ScienceNews}. This encouraged many others
 to investigate foundations of the second law, see the
proceedings of conference `Quantum Limits to the Second Law',
 San Diego, July 2002 ~\cite{QL2L}.
Many details on the thermodynamics of the model were presented in ref.
~\cite{NAlinw}. 

A thermodynamic analysis of the somewhat related spin-boson model
appeared in ref. ~\cite{ANNMR}.

\section{Quantum Brownian motion}

The Hamiltonian for a harmonic oscillator in a bath of harmonic oscillators
reads (``Caldeira-Leggett model'') 
${\cal H}={\cal H}_S+{\cal H}_B+{\cal H}_{\rm I}$, with ~\cite{NAlinw}
\BEA {\cal H}=\frac{p^2}{2m}+\half a x^2+\sum_{i}\left [
\frac{p_i^2}{2m_i}+\frac{m_i\omega_i^2x_i^2}{2}\right]
+\sum_i\left [- x_ix +\frac{c^2_i}{2m_i\omega_i^2}x^2 \right]
\EEA
The interaction Hamiltonian includes a self-interaction. 
In certain situations it is self-generated, else the prefactor
of the total $x^2$-term should be large enough to make the 
splitting with a positive $a$ is possible. 
Examples of quantum Brownian motion are:
fluctuation effects in Josephson junctions,
 low-temperature quantum transport, and
 quantum-optical systems (e.g. an ion in a Penning trap)

We assume for simplicity that the levels of the bath are equally spaced,
 $\omega _i=i\,\Delta$,  and a very small level spacing $\Delta$
corresponds to a large, extensive bath, allowing a sharp definition 
of temperature, thus providing a basis for thermodynamics.
The so-called spectral density $J( \omega)=\sum_i
\frac{\pi c_i^2}{2m_i\omega_i}\delta(\omega_i- \omega)$
is called Ohmic when $J( \omega)=\gam \omega$ for small $\omega$; 
we shall consider the quasi-Ohmic Drude-Ullersma spectrum,
where the large-$\omega$ behavior is cut-off at the Debye frequency 
$\Gamma$, i.e. $J( \omega)=\gamma \omega \Gam^2/(\omega^2+\Gam^2)$.

\subsubsection{ Stationary distribution.} 
The Wigner function of the subsystem is long known.
 By considering a Gibbsian for the
total system and summing out the bath one gets 
\BEA W(x,p)=\frac{1}{Z_x}\,
\exp(-\frac{a x^2}{2 {T}_{ x}})
\,\times\,\frac{1}{Z_p}
\exp (-\frac{  p^2}{2m {T_{  p}}}) 
\EEA
Here $T_p$ and $T_x$ are effective temperatures 
expressible in $\psi(z)=\Gamma'(z)/\Gamma(z)$,
\BEA T_x&=&T+\frac{\hbar a}{\pi m}\,\left\{
\frac{\omega_1-\Gam}{(\omega_2-\omega_1)(\omega_3-\omega_1)}
\psi\left(\frac{\beta\hbar \omega_1}{2\pi}\right)
+{\rm cyclic\,} \right\}\nn\\
T_p\!&=&\!T_x+\frac{\hbar\gam\Gam }{\pi m}\,\left\{
\frac{\omega_1}{(\omega_2-\omega_1)(\omega_3-\omega_1)}
\psi\left(\frac{\beta\hbar \omega_1}{2\pi}\right)
+{{\rm cyclic}}\right\}\nn\\\EEA
where $\omega_{1,2,3}$ are the roots of 
$\omega^3-\Gamma \omega^2+({a+\gamma\Gamma})\omega/m
-{a\Gamma}/{m}=0$ and where 'cyclic' refers to the two other
terms generated by the cyclic interchange of the $\omega_i$ ($i=1,2,3$).
In Figure 2.1 we present $T_x$ and $T_p$ as function of $T$
for a case of underdamping (left figure)
and overdamping (right figure).

\begin{figure*}
 \includegraphics [width=14cm, height=5cm] {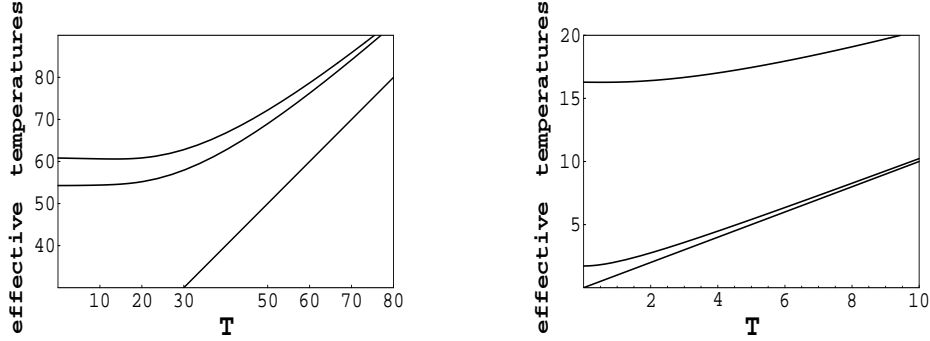}
\caption{The effective temperatures $T_p$ (upper plot), 
$T_x$ (middle plot) versus the bath temperature $T$ (lower plot) 
for $\hbar\gamma/(4\pi m) =1$, $\hbar\Gamma/(2\pi) =100$.
Left: $am/\gamma^2=80 $ (underdamping);  Right:
$am/\gamma^2=0.2$ (moderate overdamping).}
\label{fig-1}
\end{figure*}

For a weakly  coupled bath ($\gamma$ small) one gets from this:
$T_p=T_x=\half\hbar\omega_0\coth (\hbar\omega_0/2kT)$
with $\omega_0=\sqrt{a/m}$,
the standard result obtained by assuming the Gibbs distribution 
for the particle. In the absence of a bath (i.e., first $\gamma\to 0$,
then $T\to 0$) one would have $T_p=T_x=\half\hbar\omega_0$.

When the coupling between subsystem bath (characterized by $\gamma$)
is non-negligible, one has $T_p>T_x>0$, even  at $T=0$. 
The state (2.1) is non-Gibbsian since $ {T}_p {\neq} {  T}_x$,
which endangers (near-) equilibrium thermodynamics.

\subsubsection {Thermodynamic description for 
adiabatic changes}

The energy of the subsystem can be expressed through the Wigner function
\BEQ U=\langle { \H_\RS}\rangle=
\int \d p\d x {\H_\RS}(p,x)W(p,x)\equiv\int\d p\d x { \H_\RS} W \EEQ
where $ \H_\RS(p,x)=p^2/2m+ax^2/2$. For small changes in the 
(effective) mass $m$ and/or the spring constant $a$,
this allows to define the first law as 
\BEQ \d {U}=\int\d p\d x { \H_\RS} \d W+\int\d p\d x W \d { \H_\RS} 
\equiv \dbarrm { Q}+\dbarrm {{\cal W}} \EEQ
representing the heat and the work added to the subsystem, 
respectively. The latter is equal to the work extracted from 
the total system (subsystem+bath)~\cite{NAlinw}.

{\it Violation of the  Clausius inequality.} The relation
$\dbarrm Q \le T~{\rm d} S$ can now be tested at $T=0$, 
where $S$ is not needed. One finds for a change in $m$
\BEQ \dbarrm Q(T\to 0)=(\frac{\p T_p}{\p m}+\frac{\p T_x}{\p m}
+\frac{T_p}{m})\frac{\d m}{2}=
\frac{\hbar \gam}{2\pi m^2}{\rm d}m \neq 0.\EEQ
For $\d m>0$ energy is supplied by the bath to the subsystem, 
even though $T=0$. It can be shown that this energy comes 
fully from the change in interaction energy of the cloud
of bath modes around the subsystem
~\cite{NAlinw}.

A setup to test this violation in nanoscale electric circuits 
has been proposed; notice that the parameter dependence
of $T_x$ was already confirmed~\cite{ANrlc}.

{\it Violation of the  Landauer bound.} For the erasure of one bit
of information Landauer argued that an amount of heat 
$|\dbarrm Q|\ge kT\ln 2$ should be dispersed. This is a
special form of the Clausius inequality in the regime
$\dbarrm Q<0$, $\d S=-\ln 2$. However, an explicit
example for small $T$ shows that, in a process where work is 
added to the system, one can adsorb heat, $\dbarrm Q>0$, and 
nevertheless become more localized, ${\rm d} S<0$~\cite{ANLand}.

\subsubsection {Non-adiabatic energy dispersion}

The work can be decomposed in the adiabatic part and the 
energy dispersion, 
$ \dbarrm  {\cal W}=\dbarrm {\cal W}_{\rm  {adiabatic}}+\dbarrm  \Pi$.
For a small, smooth and slow change of the spring constant,
$a(t)=[1+\alpha(t)]a$, one gets at small $T$ to order $\alpha^2$ 
\cite{NAlinw} \BEQ\label{daalpddalp}
\frac{\dbarrm  \Pi}{\d t}= T^2\,\dot\alpha^2+\dot\alpha\ddot \alpha
-\dot\alpha\partial_t^3\alpha
\EEQ
where numerical factors have been left out.
For a completed change the adiabatic term drops out and the dispersion
brings $\Delta {\cal W}=
\Delta \Pi=\int_{-\infty}^\infty\d t\,\dbarrm  \Pi/\d t
=\int_{-\infty}^\infty\d t (T^2\,\dot\alpha^2+\ddot\alpha^2)> 0$,
confirming the Thomson formulation of the second law (cycles cost work)
~\cite{ANthomson}.

When $T$ is small and $\alpha$ is a slow function of time,
the $\dot\alpha\ddot\alpha$ term dominates;
only in its	 full integral its positive and negative parts cancel. 
Thus a negative $\frac{\dbarrm  \Pi}{\d t}$ is possible. 
This poses a counterexample to the formulation of the second law 
stating that the rate of energy dispersion be positive~\cite{KondPrig}.

The fact that at low $T$ the $\dot\alpha\ddot\alpha$-term is leading can 
be exploited. Cycles in $a(t)$ can be constructed 
in which energy is extracted. In figure 2.2 we show bell-shaped curves
for $\alpha(t)=\alpha_{\rm max}h(t)$, which become slower and slower.
That construction is needed to let the $\dot\alpha\ddot \alpha$-term
overcome the inherent loss due to the first and third term~\cite{NAlinw}.

\begin{figure*}
\vspace{-2cm}
\includegraphics[width=15cm,height=7cm]{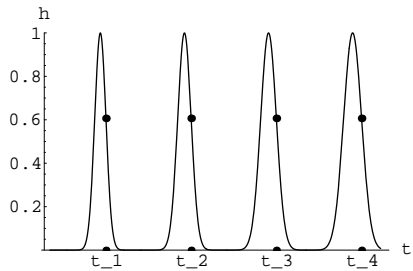}
\caption{ Schematic plot of the cyclic changes in the spring constant,
where successive cycles are slower and slower.
$h$ characterizes the size of the change and $t$ denotes
the dimensionless time. The interval $-\infty<t<t_1$
marks the process that establishes the nonequilibrium state at
$t=t_1$. The picture shows three full cycles,
between the bullets.}
\label{fig:perpcycles}
\end{figure*}


The total dispersion is positive, because more work is needed to create 
the non-equilibrium state at $t_1$ than can be recovered in the cycles.
Nevertheless this can be called a ``Perpetuum mobile of second kind''
~\cite{NAlinw}. The maximal number of work extraction cycles is of 
order ${1}/{T}$. Optimal extraction occurs, however, by making one 
cycle, due to the trend for dispersion in completed cycles.

\subsubsection {Entropic considerations}

Entropy is historically a newer concept than heat and work.
It has come as a surprise that in the present field 
entropy arguments fail even quicker.
With the Clausius inequality being broken, there is no sense in
defining a ``thermodynamic'' entropy 
$S_{\rm thermo}=\int_0^T\d T' (\dbarrm Q/\d T')/T'$; this quantity
would not have a statistical interpretation, and, actually, 
the integral would not even converge at its lower bound.

The most common formulation of the second law is: 
the (coarse grained) entropy of a closed system cannot decrease. 
It becomes for an open subsystem
(immersed in its bath): the rate of entropy production cannot be
negative. By defining the flow of Boltzmann entropy, we came to
the conclusion that the rate of Boltzmann entropy production 
can be negative already for moderate temperatures~\cite{NAlinw}.

\section{Bath generated work extraction in two-level systems} 

Another application is the spin-boson model, often used in NMR 
and ESR physics, quantum optics and spintronics. 
Its Hamiltonian reads ~\cite{ANNMR}
\BEA
\label{ham}
\H&=&\H(\Delta)=\H_S+\H_B+\H_I,\quad \\ \H_S&=&
\frac{\eps}{2}\,\si _z+\frac{\Delta(t)}{2}\si_x, \quad
\H_B=\sum _k\hbar\om _k\ha^{\dagger}_k\ha _k, \quad
\H_I=\frac{1}{2}\sum _kg_k(\ha _k^{\dagger}+\ha _k)\si _z. \nn
\EEA
This is a spin-$\frac{1}{2}$ interacting with a bath of harmonic
oscillators. $\si _x$, $\si _y$ and $\si_z=-i\si_x\si_y$ are Pauli 
matrices, and $\ha _k^{\dagger}$ and $\ha_k$ are the creation and 
annihilation operators of the bath oscillator with the index $k$, 
while the $g_k$ are the coupling constants.  
For an electron in a magnetic field $B$, $\eps= \bar g\mu_B B$ is the
energy, with $\bar g$ the gyro-magnetic factor and $\mu_B$ the Bohr
magneton.  We take the transversal field $\Delta(t)=0$ except during pulses.
This model is a prototype of a variety of physical systems \cite{rmp}, and
known to be exactly solvable \cite{rmp,lu}, since the $z$-component of
the spin is conserved, and with it the spin energy.  Physically it
means that we restrict ourselves to times much less than $\T_1$
(relaxation time of $\si_z$).  In
ESR physics~\cite{nmr} the model represents an electron-spin
interacting with a bath of phonons, for NMR it can represent a nuclear
spin interacting with a spin bath, since in certain natural limits the
latter can be mapped to the oscillator bath.  In quantum optics it is
suitable for describing a two-level atom interacting with a photonic
bath \cite{opt}.

One typically takes a quasi-Ohmic spectral density \cite{rmp} 
$J(\omega)=\sum_k g_k^2\delta(\omega_k-\omega)/(\hbar\omega_k)
=g\,\,\hbar\, \exp(-\om /\Gamma)/\pi$,
where $g$ is a dimensionless damping constant and the 
exponential cuts off the coupling at $\omega\gg \Gamma$, 
the maximal frequency of the bath. 
Since $\Delta=0$, one has conservation of $\si_z(t)=\si_z(0)$ 

{\it Separated initial state}.
To describe situations, where the spin was suddenly brought into the
contact with the bath, e.g. an electron injected into semiconductor,
atom injected into a cavity, or exciton created by external radiation,
we make the assumption that initially, at 
$t=0$, the spin and the bath are in a separated state,
the latter being Gibbsian, $\rho (0)=\rho _{S}(0)\otimes 
\exp (-\beta \H_B)/Z_B$,
where $\rho _S(0)$ is the initial density matrix of the spin.
In terms of the Larmor frequency $\omega_0=\eps/\hbar$
this brings for the evolution of the transverse components
\BEA
\label{k7}
\langle \si_x (t)\rangle &=&
[\cos\om t\langle \si_x (0)\rangle-\sin\om t\langle\si_y (0)\rangle]
\,e^{-t/\T_2},\nn\\
\langle \si_y (t)\rangle &=&
[\cos\om t\langle \si_y (0)\rangle+\sin\om t\langle\si_x (0)\rangle]
\,e^{-t/\T_2},\EEA
where $\T_2=\hbar/(gT)$ is the transversal decay time.
 
{\it The von Neumann entropy} equals
$S=-\tr \rho_S\ln\rho_S=-p_1\ln p_1-p_2\ln p_2$, where $
p_{1,2}=\half\pm\half|\langle{\vec{\sigma}}\rangle|$.
The decay of $\langle{\sigma_{x,y}}(t)\rangle$
makes  the von Neumann entropy increase.
Since there is no heat flow - the energy is conserved - this is
in agreement with a formulation of the second law: the entropy 
of a closed system, or of an open system without energy transfer, 
cannot decrease.

{\it A sudden pulse}. 
A fast rotation around the $x$-axis is described by
taking $\Delta\neq 0$ during a short time ~\cite{nmr}, yielding  
$(\si_{y\,,z})'=\si_{y,z}\cos \theta\pm\si _{z,y}\sin \theta$.
During the sudden switchings of $\Delta(t)$ 
the density matrix is just rotated.

Our main interest is work extraction from the bath.
In order to ensure that the pulse does not change the energy of the spin, 
we first consider the case $\eps =0$, where the spin has no energy. 
For small $g$, $\theta=-\pi/2$ and $ t\gg 1/\Gamma$ 
the work appears to be
\BEQ\label{W1==}
 W_1=\frac{g\,\,\hbar\,\Gamma}{2\pi}+
\frac{gT}{2}\langle\si_x(0)\rangle\,e^{-t/\T_2}\EEQ 
If for a fixed $t$, temperature is neither too large nor too small,
$Te^{-t/\T_2}> \,\,\hbar\,\Gamma/\pi$, 
work can be extracted ($W_1<0$),  provided
the spin started in a coherent state  $\langle\si_x(0)\rangle= -1$. 
This possibility to {\it extract work} from the equilibrated bath
($t\gg 1/\Gamma$) contradicts the Thomson's formulation of the
second law out of equilibrium. It
disappears on  the timescale $\T_2$, because then the spin looses its
coherence, $\langle\si_{x,y}(t)\rangle\to 0$. 
Classical variations ($\pm\pi$ pulses, which do not involve coherence) 
can extract work only from a non-thermalized bath, i.e. 
for times $\sim 1/\Gamma$. Thus, the effect is essentially quantum
mechanical.

{\it Initial preparation via a rotation.}
Starting from a Gibbsian of the total system,
at $t=0$ the spin is rotated 
(``zeroth pulse'') over an angle $-\half\pi$ around the $y$-axis.
This maps $\si_x\to\si_z$, $\si_z\to-\si_x$.
Such a state models the optical excitation of the spin, as done
in NMR and spintronics. The problem remains exactly solvable.
Taking $\theta=-\half\pi$ in the first pulse one now gets 
\BEA W_1\approx\frac{g\,\,\hbar\,\Gamma}{2\pi}-
\left[\frac{\eps}{2}\sin\omega_0t+\frac{gT}{2}\cos\omega_0t\right]
\tanh\frac{\beta\eps}{2}\,\,e^{-t/\T_2}
\EEA
The work decomposes as $ W_1=\Delta U-\Delta Q$, with
\BEA \Delta Q\approx\frac{g}{2}
\left[-\frac{\,\,\hbar\,\Gamma}{\pi}+T\cos\omt\,\tanh\frac{\beta\eps}{2}
\,e^{-t/\T_2}\right],
\EEA
quite similar to $-W_1$ of Eq. (\ref{W1==}).
An interesting case is where work is performed by the total
system ($W_1<0$) solely due to  heat from the bath ($\Delta Q>0$,
$\Delta U=0$). This process, possible by choosing 
$t\approx 2\pi n/\omega_0$ with integer $n$, 
involves a $\Delta Q>0$ , $\Delta S=0$,
which violates the Clausius inequality.

The work needed at $t=0$ to rotate the spin is
$W_0=(\eps/2)\,\tanh[\beta\eps/2]+g\,\,\hbar\,\Gamma/(2\pi)$.
The extracted work is smaller, confirming Thomson's equilibrium
formulation for cyclic changes 
($\Delta=0$ before and after the pulses) ~\cite{ANthomson}.

{\it Two pulses in a rotated initial Gibbsian state.}
If there are many (very weakly interacting) spins, 
each in a slightly different external field, 
an inhomogeneous broadening of the $\omega_0=\eps/\hbar$ 
line occurs, for which we  assume the distribution 
\BEQ p(\omega_0)=
\frac{2}{\pi} \frac{[\T_2^\ast]^{-1}}{(\omega_0-\bar\omega_0)^2
+[\T_2^\ast]^{-2}}
\EEQ
with average $\bar\omega_0$ and inverse width $\T_2^\ast$,
typically much smaller than $ \T_2$.
In this case the gain for a single pulse is washed out, 
leaving only the loss $\Delta Q=-g\,\,\hbar\,\Gamma/2\pi$,
so two pulses are needed.
We consider again the rotated initial Gibbsian state,
and perform a first $-\half\pi$ pulse around the $x$-axis
at time $t_1$ and a second $\half\pi$ pulse 
at time $t_2$ (the standard $\half\pi$, $\pi$ 
combination would not expose an interesting role of the bath).
For the total work $W=W_1+W_2$ the averaging over $\omega_0$ brings
\BEA W&=&\frac{g\,\,\hbar\,\Gamma}{\pi} 
-\frac{\hbar}{4} e^{-t_2/\T_2}e^{-|\Delta t|/\T_2^\ast}
\tanh\frac{\beta\,\,\hbar\,\bar\omega_0}{2}\times \\
&&\left
\{\bar\omega_0\sin\bar\omega_0\Delta t\right. 
 +
[\frac{1}{\T_2}-\frac{{\rm sg}(\Delta t)}{\T_2^\ast}
(1+\frac{\beta\,\hbar\bar\omega_0}{\sinh\beta\,\hbar\bar\omega_0})]
\cos\bar\omega_0\Delta t\,\} \nn \EEA
For  $\Delta t\equiv t_2-2t_1$ near $2\pi n/\bar\omega_0$ 
such that the odd terms 
cancel, this again exhibits work extracted solely from the bath.

{\it Feasibility.}
Let us notice that 
work and heat were measured in NMR experiments more than 35 years ago
\cite{Schmidt} and this technique (glue the sample on a copper wire and
measure the change in its resistance)
can be used to test the violation of the Clausius inequality.
The ongoing activity for implementation of quantum computers
provides experimentally realized examples of two-level systems, which 
have sufficiently long ${\cal T}_2$ times,
and admit external variations on times smaller than ${\cal T}_2$:
({\it i}) for atoms in optical traps ${\cal T}_2\sim 1$s, 
$1/\Gamma\sim 10^{-8}$s, and there are efficient methods for
creating non-equilibrium initial states and manipulating atoms by
external laser pulses \cite{atoms}; ({\it ii}) for an
electronic spin injected or optically excited in a semiconductor
${\cal T}_2\sim 1\,\mu$s \cite{spintronics}; 
({\it iii}) for an exciton created in a quantum dot  
${\cal T}_2\sim 10^{-9}$s \cite{exciton} (in cases ({\it ii}) and ({\it iii})
$1/\Gamma\sim 10^{-13}$s and femtosecond ($10^{-15}$s) laser pulses are
available); ({\it iv}) in NMR physics ${\cal T}_2\sim 10^{-6}-1$ s 
and the duration of pulses can be comparable with 
$1/\Gamma\sim 1 \,\mu$s. 

\subsubsection{ Bath-induced gain without inversion} 

A two-level system with population inversion, i.e. with a negative 
temperature, is a working mechanism of lasers and masers. 
In this context a bath is
typically considered as a source of undesirable noises and 
relaxation towards equilibrium~ \cite{opt}. 
The bath can nevertheless play a totally 
different role, namely in {\it assisting} work extraction (gain) by means 
of a {\it positive} temperature state in the two-level system. 
In absence of coupling to the bath such an effect is strictly
prohibited by the second law applied to a  positive temperature 
spin state \cite{ANthomson}.

We consider separated initial conditions
with $\langle \si_x(t)\rangle=\langle \si_y(t)\rangle=0$,
and apply a $-\half\pi$ pulse around the $x$-axis at time $t_0=0^+$, and a 
$\half\pi$ pulse at $t$. 
For $t\gg 1/\Gamma$ the work $W=\Delta U-\Delta  Q$ is set by:
\BEA
&&\Delta U=-\frac{\eps}{2}\,[1-\,e^{-t/\T_2}\cos\omega_0t\,]
\langle\si_z\rangle
+\frac{g\eps}{4}\,e^{-t/\T_2} \sin\omega_0t\nn\\ 
&&\Delta  Q=-\frac{g\hbar\Gamma}{\pi}+
\half gT\,e^{-t/\T_2}\sin\omega_0 t\langle\si_z\rangle
\EEA
In the inversion-free case the initial state of the spin 
is a Gibbsian connected with temperature $T_0=1/\beta_0$, for which
$\langle\si_z\rangle=-\tanh\half\beta_0\eps\le 0$.
For $T_0=\infty$ one has the completely random state, $\langle
\si_{x,y,z}\rangle=0$.  The work $W$ can be negative (gain) provided
$\eps> 4\hbar\Gamma/\pi$. This situation can be met in quantum optical
two-level systems \cite{opt,atoms} and in NMR \cite{nmr1}.  This
mechanism concerns work extraction {\it with help of the bath} (it
disappears for $g\to 0$), but {\it not from the bath}, since now
$\Delta  Q<0$.  The origin of the effect is that although the state of
the spin was completely disordered initially, the first pulse does
generate some coherence.  Due to back-reaction of the bath one has
after the pulses $\langle\si_y(t)\rangle=\half g
\exp(-t/\T_2)\sin\omega_0 t$.

At finite $T_0$ the term $\Delta U$
can still be negative when $T_0>\eps/g$, which can be met for
not-too-small $g$, a condition anyhow needed for having a sizeable
effect.  From a thermodynamic point of view the gain can just be seen
as a flow of energy from a high temperature (of the spin) to a lower
one (of the bath), and the outside world (gain).

There exist other mechanisms for inversionless gain \cite{opt}; 
the crucial difference is that they operate with (at least) three-level 
systems (atoms), and | most importantly |
the effect appears due to special, non-thermal states of the atom
itself. Frequently these states involve a hidden inversion \cite{opt}.

\section{Exact theorems}

There are a few exact theorems that help to reach a final 
formulation of the second law.
They all start from initial Gibbsian equilibrium.

\begin{itemize}

\item[1] The Thomson formulation (cycles cost work) 
allows an exact proof when starting in 
equilibrium~\cite{ANthomson}.

\item[2] The amount of heat that can go from an equilibrium low 
temperature bath to an equilibrium high temperature bath is limited
 and proportional to the interaction energy. In particular, this 
forbids steady energy currents from low to high $T$~\cite{ABNsd}

\end{itemize}

In previously discussed models these theorems can be checked whenever 
they apply, that is to say, when the initial state is Gibbsian.

\section{What went wrong with standard thermodynamics ?}

The present research considers small quantum 
systems not so weakly coupled to their baths, that
the interaction energy can be neglected.

For such systems a number of complications show up in the 
formulation of thermodynamics. We have indeed shown counterexamples
against various formulations, in particular those involving the entropy.
Measurements have been proposed to test the
most obvious one, the breakdown of the Clausius inequality, e.g.
for pulses  in NMR-physics.
This immediately implies the possibility to
violate the Landauer bound for erasure of information.
Further unexpected aspects are a negative rate of energy dispersion and
the possibility of work extraction cycles.

On the other hand, there are exact theorems which save some formulations
(mostly involving only energy or work) provided the initial state
is Gibbsian and the work source is not correlated to the test system
and its bath.

The aim of the present field is to establish thermodynamically 
unexpected energy flows on  microscopic or mesocopic scales. 
The field of laser physics is a promising play ground for such phenomena.

\end{document}